\documentclass[10pt,conference]{IEEEtran}
\IEEEoverridecommandlockouts
\usepackage{cite}
\usepackage{amsmath,amssymb,amsfonts}
\usepackage{algorithmic}
\usepackage{graphicx}
\usepackage{textcomp}
\usepackage{siunitx}
\usepackage{xspace}
\usepackage{xcolor}
\def\BibTeX{{\rm B\kern-.05em{\sc i\kern-.025em b}\kern-.08em
    T\kern-.1667em\lower.7ex\hbox{E}\kern-.125emX}}
\usepackage{cancel}
\usepackage{soul}
\usepackage{enumitem}
\usepackage{minted}
\usepackage{url}
\usepackage{tikz}
\usepackage{lstlinebgrd}
\usepackage[hidelinks]{hyperref}
\usepackage{multirow}
\usepackage{tabularx}
\usepackage{array}

\usepackage{multicol}
\usepackage{threeparttable}
\usepackage{booktabs}
\usepackage{listings}

\usepackage{svg}
\renewcommand{\arraystretch}{1.2}

\usepackage{tcolorbox}%
\definecolor{mycolor}{rgb}{0.122, 0.435, 0.698}% Rule colour
\definecolor{gray1}{gray}{0.3}

\definecolor{darkgreen}{rgb}{0.0, 0.5, 0.0}
\definecolor{darkred}{rgb}{0.82, 0.1, 0.26}

\newcommand{\result}[1]{%
\begin{tcolorbox}[colframe=mycolor,boxrule=0.5pt,arc=4pt,
      left=6pt,right=6pt,top=6pt,bottom=6pt,boxsep=0pt,width=\columnwidth]%
      {#1}
\end{tcolorbox}%
}
\newcommand{\method}{\textsc{CodeZoom}\xspace}
\newcommand{\cc}{Claude Code\xspace}
\newcommand{\todo}[1]{}
\renewcommand{\todo}[1]{{\color{red} TODO: {#1}}}
\newcommand{\add}[1]{}
\renewcommand{\add}[1]{{#1}}
\newcommand{\del}[1]{}
\renewcommand{\del}[1]{{\color{red} \sout{#1}}}
\newcommand{\etal}{\textit{et al}.\xspace}
\newcommand{\ie}{\textit{i}.\textit{e}.\xspace}
\newcommand{\eg}{\textit{e}.\textit{g}.\xspace}

\newcommand{\foldersymbol}{\raisebox{-0.3\height}{\includegraphics[height=0.25cm]{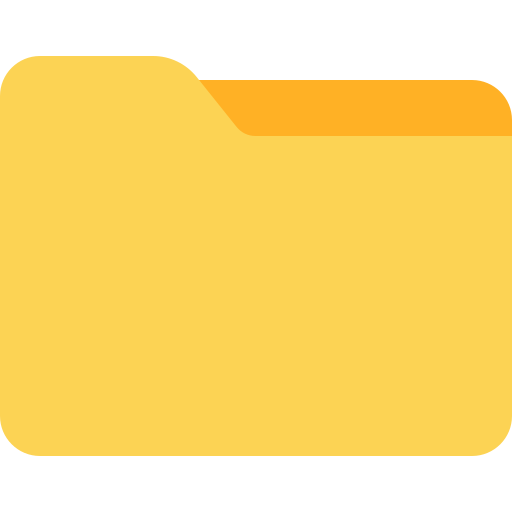}}}
\newcommand{\documentsymbol}{\raisebox{-0.3\height}{\includegraphics[height=0.25cm]{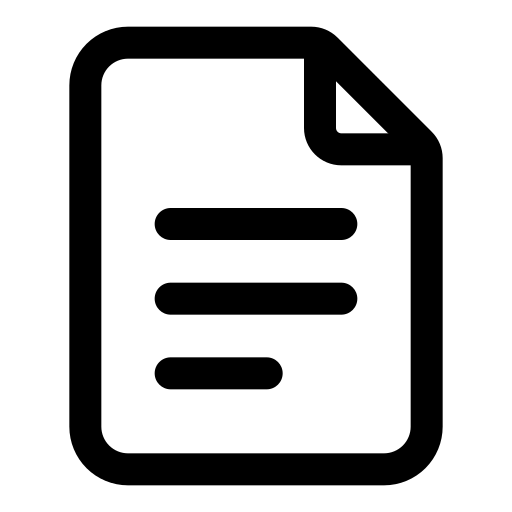}}}
\newcommand{\codesymbol}{\raisebox{-0.3\height}{\includegraphics[height=0.25cm]{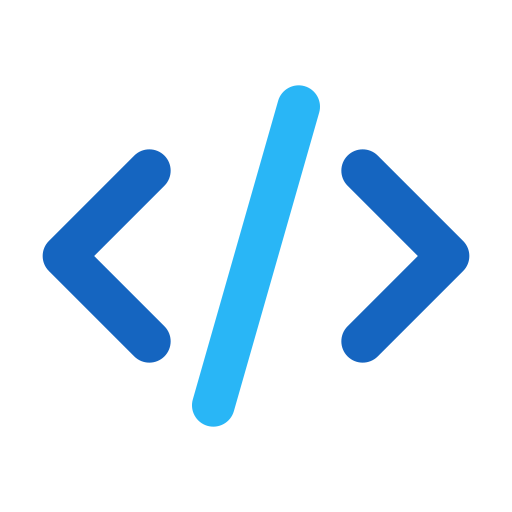}}}
\newcommand{\pseusymbol}{\raisebox{-0.3\height}{\includegraphics[height=0.25cm]{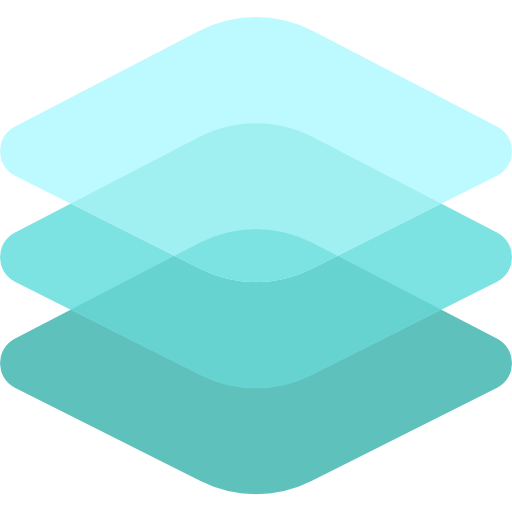}}}
\newcommand{\navigationsymbol}{\raisebox{-0.3\height}{\includegraphics[height=0.25cm]{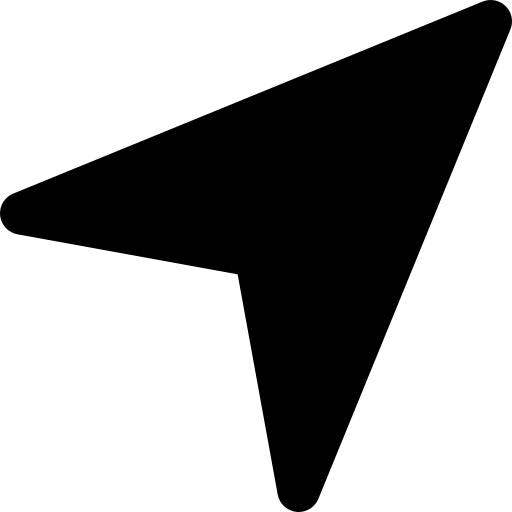}}}

\begin{document}

\title{Code Semantic Zooming}
%\author{
%\IEEEauthorblockN{Anonymous Author(s)}
%}

\author{\IEEEauthorblockN{Jinsheng Ba}
\IEEEauthorblockA{ETH Zurich}
\and
\IEEEauthorblockN{Sverrir Thorgeirsson}
\IEEEauthorblockA{ETH Zurich}
\and
\IEEEauthorblockN{Zhendong Su}
\IEEEauthorblockA{ETH Zurich}
}

\maketitle

\begin{abstract}
Recent advances in Large Language Models (LLMs) have introduced a new paradigm for software development, where source code is generated from natural language prompts. While this paradigm significantly boosts development productivity, building complex, real-world software systems remains challenging because natural language offers limited control over the code generation process. Inspired by the historical evolution of programming languages toward higher levels of abstraction, we advocate for a high-level abstraction language that gives developers greater control over LLM-assisted code writing. To this end, we propose Code Semantic Zooming (CodeZoom), a novel approach based on pseudocode that allows developers to iteratively explore, understand, and refine code across multiple layers of semantic abstraction. In a within-subjects user study ($n{=}26$), our method matches a state-of-the-art coding agent, Claude Code, on usability while producing a large effect on code comprehension: over 90\% of participants reported feeling more in control of design decisions when using CodeZoom compared to using Claude Code. On objective multiple-choice questions about participants’ own implementations, an exploratory measure added mid-study, a subset of participants ($n{=}16$) answered 56\% of comprehension questions correctly with CodeZoom, compared with 9\% with Claude Code (Wilcoxon $p=.005$).
\end{abstract}


\section{Introduction}

In recent years, the paradigm of code writing has undergone a substantial shift driven by Large Language Models (LLMs)\cite{zhao2023survey, chen2021evaluating, liu2024large}. Natural language, most commonly English, has effectively become a new programming language. Developers provide specifications or requirements in natural language prompts, and LLMs generate the corresponding source code. This transformation reduces manual coding effort, allowing developers to focus on higher-level tasks such as design and problem-solving. The approach was termed \emph{vibe coding}~\cite{vibecoding} and has been rapidly integrated into modern coding tools, including Codex\cite{codex}, Cursor~\cite{cursor}, Gemini CLI~\cite{gemini}, Claude Code~\cite{claudecode}, and Copilot~\cite{copilot}.

However, natural language offers insufficient control over the code generation process, making it difficult to develop sophisticated real-world software projects using vibe coding alone. The first challenge lies in intention expression. Unlike traditional programming languages such as Python or C, which provide precise syntax and semantics, natural language is inherently ambiguous, underspecified, and context-dependent. This ambiguity often leads to misinterpretations of requirements, resulting in code that diverges from the developer’s true intent.  
The second challenge is code comprehension. Although LLMs can generate syntactically correct and seemingly plausible code, there is no guarantee of semantic correctness or alignment with user intent in natural language. While LLM-generated summaries can provide partial assistance, fully understanding and refining the underlying program logic remains difficult. Developers must still rely on testing or manual inspection to ensure that the generated code satisfies functional requirements.
These difficulties are amplified by the iterative nature of software development, where code is continuously revised to add or modify features. Vibe coding offers little control over code generation in such settings. Even if each iteration achieves a correctness rate of 95\%, repeated rounds of generation will compound errors, ultimately reducing overall correctness to an unacceptable level.

\begin{figure}
    \centering
    \includegraphics[width=\linewidth]{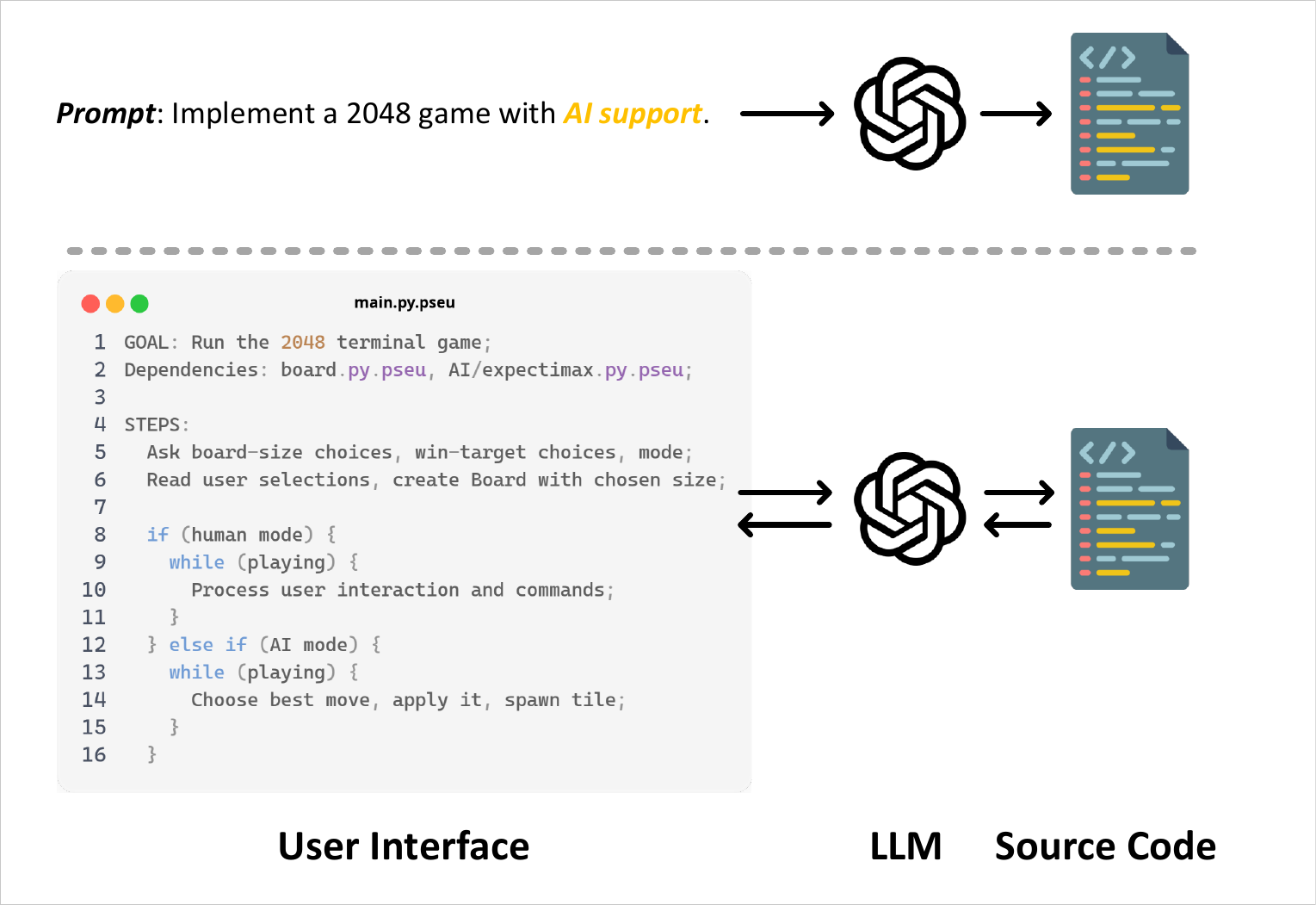}
    \caption{Existing vibe-coding and \method to implement a 2048 game.}
    \label{fig:motivation}
\vspace{-5mm}
\end{figure}

We envision the need for a high-level abstract language to provide greater controllability in LLM-assisted code writing. Such a language must remain as intuitive as natural language while encoding the structural scaffolding necessary to guide and constrain code generation. \add{While various formalisms could serve this purpose, we instantiate this concept using pseudocode. Because pseudocode is ubiquitous, modern LLMs already possess a robust, pre-trained understanding of its semantics.}
Pseudocode~\cite{roy2006designing} offers a high-level, human-readable description of program logic that blends natural language with formal programming constructs. Leveraging this dual accessibility, we establish a bidirectional translation mechanism between executable source code and its pseudocode.
However, traditional pseudocode is inherently static and cannot capture code semantics at varying levels of granularity. To address this limitation, we propose \emph{Code Semantic Zooming} (\method), a grammar-driven approach that leverages LLMs to dynamically adjust the abstraction level of pseudocode. This approach makes the generated code more controllable, offering both human interpretability and machine guidance.

\autoref{fig:motivation} presents a motivating example of comparing a pure vibe coding tool with \method for writing code. While an LLM can generate standard games from pre-trained knowledge alone, the limitations of pure natural language prompts become critical when addressing novel feature requests where pre-trained knowledge falls short. 

%The task is to implement a 2048 game, a sliding tile puzzle in which players combine numbered tiles by moving them in four directions to create a tile with the number 2048.

As shown in the upper half of \autoref{fig:motivation}, users write natural-language prompts directly to generate source code. One might provide the prompt ``\emph{Implement a 2048 game with AI support}''. However, this prompt can be interpreted in multiple ways, such as whether the AI should fully automate gameplay or merely provide hints to human players. Such distinctions are difficult to express unambiguously in natural language by humans. 
%To inspect the generated code, we might extend the prompt with ``\emph{Summarize the generated code}'', obtaining a response such as: ``\emph{I implemented a self-contained 2048 game in Python with optional AI control. The code includes the game logic, a command-line interface for human play, and an AI using an expectimax-like search with a heuristic evaluation}''. This summary does not clarify how or where the AI algorithm is integrated into the code.
In contrast, as shown in the lower half of \autoref{fig:motivation}, \method provides an interface in which users write and edit pseudocode, which LLMs then translate into source code. Operating on pseudocode rather than natural language reduces ambiguity by allowing explicit specification of structural controls such as \lstinline{if} and \lstinline{else}. For example, users can clearly indicate that the AI should perform fully automatic gameplay, as a mode distinct from human play. For code validation, \method translates the generated source code back into pseudocode, making it explicit that the AI operates as a separate play mode rather than merely providing next-step hints. In addition, the abstraction level of the pseudocode can be adjusted to be more fine- or coarse-grained, enabling users to inspect details at the desired granularity.

We implemented \method as a set of skills, which are pre-packaged folders of task-specific instructions for LLM-based coding agents, totaling roughly 2,000 lines, and developed it on top of Claude Code. Because the skill interface is portable, the same implementation can be readily adapted to other agents such as Codex. To evaluate the approach, we conducted a within-subjects user study with 26 graduate students, each of whom completed two programming tasks: implementing a 2048 game from scratch and reproducing a real feature request in an open-source codebase. \method was rated as usable as the Claude Code baseline (SUS $69.0$ vs.\ $71.3$, $p{=}0.62$) while delivering substantial gains in code comprehension: on a 7-point Likert item, participants rated their understanding of the codebase at $5.50$ with \method versus $3.35$ with Claude Code ($p{<}0.001$, Cohen's $d{=}1.02$), and on objective multiple-choice questions about their own implementations, participants answered 56\% of comprehension questions correctly under \method versus 9\% under \cc (Wilcoxon $p=.005$), and over 90\% of participants reported feeling more in control of design decisions when using \method rather than \cc. 
%The artifact is available at \url{https://figshare.com/s/d5b4235453936168f308}.

Overall, we make the following contributions:
\begin{itemize}
\item We propose a pseudocode approach \method for LLM-assisted code writing.
\item We introduce Code Semantic Zooming, enabling dynamic control over the granularity of pseudocode.
\item A user study ($n = 26$) demonstrates the code-generation controllability gained by \method. 
\end{itemize}
\section{Background}

The evolution of programming is defined by a continuous drive toward higher abstraction, aiming to improve expressiveness and maintainability. Early machine and assembly languages offered precise hardware control but were tedious and error-prone. The advent of high-level languages like FORTRAN~\cite{backus1978history} and COBOL~\cite{sammet1978early} marked a transformative shift, abstracting machine details to democratize software creation and let developers focus on problem-solving.

As software complexity escalated, new paradigms emerged to manage it. Structured programming~\cite{dahl1972structured, dijkstra2021go} and C~\cite{ritchie1978c} introduced disciplined control flows, while declarative approaches~\cite{lloyd1994practical, date1989guide} prioritized computational goals over execution steps. Object-oriented programming (OOP)~\cite{rentsch1982object, kay1996early, stroustrup1986overview} further enabled modular, reusable architectures by encapsulating state and behavior. Today, multi-paradigm languages like Python and Rust~\cite{thaker2020python, hanus2007multi} dominate, allowing developers to seamlessly blend these styles and, aided by robust package ecosystems~\cite{decan2019empirical}, drastically accelerate productivity.

Recently, large language models (LLMs) have pushed abstraction further, elevating natural language to a primary interface. Unlike Knuth's literate programming~\cite{knuth1984literate}, which intertwined explanatory prose with human-written source code, modern ``vibe coding''~\cite{vibecoding} relies on models to translate plain-language goals directly into execution. This freedom, however, comes at a cost: natural language is inherently ambiguous and underspecified, making it hard to precisely control generated code, validate its correctness, and iteratively refine it for complex systems.

In this work, we argue that the next step in this trajectory should combine natural language with structured, semantically meaningful representations, preserving the expressiveness of natural language while reintroducing the controllability needed for robust LLM-assisted code writing.

\section{Approach}\label{sec:approach}
In this section, we introduce Code Semantic Zooming (\method), an approach that enables multiple layers of semantic abstraction of source code using pseudocode, aiming to enhance the controllability of LLM-assisted code writing.

\begin{figure}
    \centering
    \includegraphics[width=0.9\linewidth]{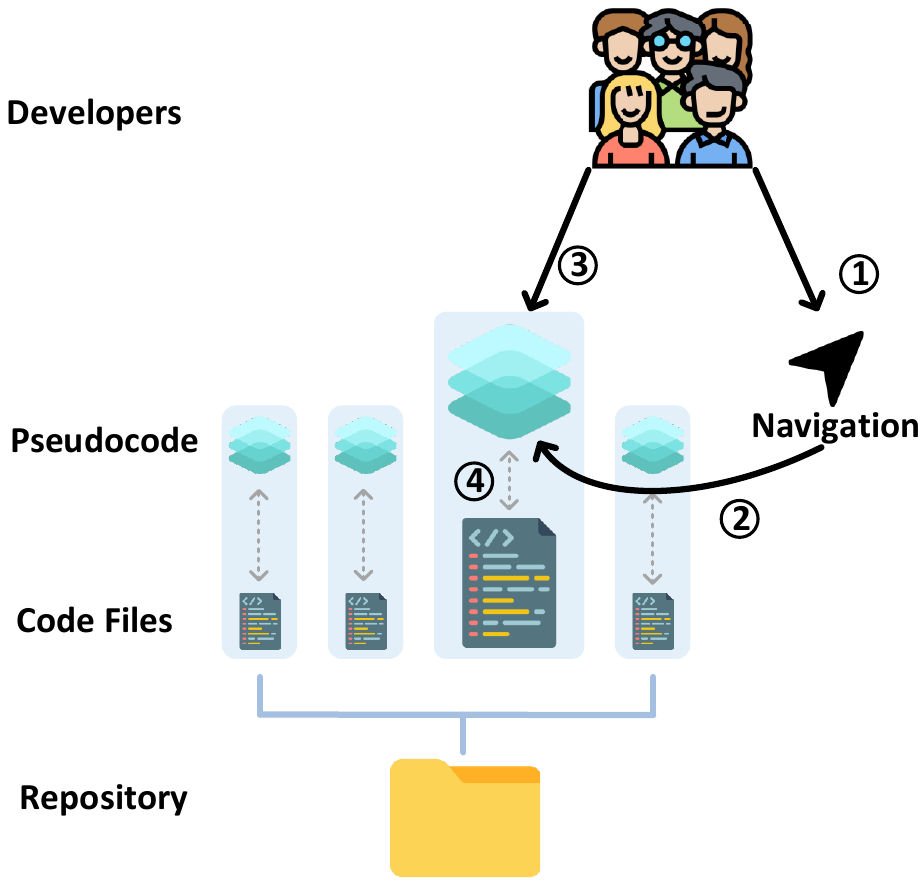}
    \caption{An overview of \method. The system abstracts each code file into a corresponding pseudocode layer. Developers utilize a navigation mechanism to locate specific pseudocode, serving as an interface to interact with and control the corresponding source code.}
    \label{fig:overview}
\end{figure}

\paragraph{Approach overview}
\autoref{fig:overview} presents an overview of \method. Suppose we are given a code repository containing four source code files. \method adds a pseudocode representation for each source code file and establishes a bidirectional translation between pseudocode and source code, enabling users to inspect and refine source code through pseudocode. In addition, \method constructs a navigation file that organizes the overall execution flow across all files. A typical use case begins with developers browsing the navigation file to understand the repository-level execution flow (\textcircled{1}). Next, in \textcircled{2}, they navigate to the specific pseudocode file of interest. Finally, in \textcircled{3}, users interact with the pseudocode by dynamically zooming in to inspect detailed logic or zooming out to abstract away low-level details for a higher-level view. This mechanism helps users quickly locate the parts of the code that require modification. Once the desired changes are expressed in pseudocode, they are propagated back to the source code, as shown in \textcircled{4}. The pseudocode is then regenerated from the revised code, allowing users to iteratively verify and refine program semantics until the implementation fully aligns with their intent.

\paragraph{Application scenarios}
\method targets two typical application scenarios across software development. 

\emph{Scenario 1: Initializing a new code repository.} When initiating a new project, developers often start with a high-level design but struggle to maintain control as the implementation grows in scale and complexity, where LLM-generated code frequently drifts away from the intended architecture. With \method, developers can first convert their high-level design into coarse-grained pseudocode that captures key logic. They then translate the pseudocode into executable source code and progressively zoom in to refine details. This top-down workflow keeps developers in charge of design decisions while delegating routine implementation to LLMs, thereby enhancing the controllability of LLM-assisted code writing.

\emph{Scenario 2: Revising an existing code repository.} When developers need to modify an unfamiliar code project, such as adding a feature or fixing a bug in an open-source project from GitHub, they typically face a steep comprehension barrier across hundreds of files. With \method, the navigation file offers an immediate high-level view of the execution flow, allowing developers to quickly locate the relevant files. They can then inspect the corresponding pseudocode at a comfortable level of abstraction, zooming in only on the regions that require modification. Edits expressed in pseudocode are propagated back to the source code, sparing developers from wading through unfamiliar low-level syntax while ensuring that changes faithfully reflect their intent.

We explain below how to navigate to specific files and interact with them when using \method.

\begin{figure}
\begin{lstlisting}[caption={An example of code repository using \method.},captionpos=t, label=lst:repo, escapeinside=&&, linebackgroundcolor={\btLstHL{4,6,8,10,12}}]
.
|-- &\documentsymbol& README.md
\-- &\foldersymbol& src
    |-- &\navigationsymbol& navigation.md
    |-- &\codesymbol& main.py
    |-- &\pseusymbol& main.py.pseu
    |-- &\codesymbol& board.py
    |-- &\pseusymbol& board.py.pseu
    \-- &\foldersymbol& AI
        |-- &\navigationsymbol& navigation.md
        |-- &\codesymbol& expectimax.py
        |-- &\pseusymbol& expectimax.py.pseu
\end{lstlisting}
\end{figure}

\subsection{Navigation to the Code Files}

\autoref{lst:repo} presents an example code repository using \method, which contains the code for a 2048 game. \method augments the repository with two categories of auxiliary files, both highlighted in grey in \autoref{lst:repo}: a ``\lstinline{navigation.md}'' file in each subfolder, which provides an overview of all code files in that directory, and a ``\lstinline{*.pseu}'' file, which contains the pseudocode abstraction of each source code file. Users can traverse the ``\lstinline{navigation.md}'' files to locate the specific pseudocode files they are interested in. 

\begin{figure}
    \centering
    \includegraphics[width=0.95\linewidth]{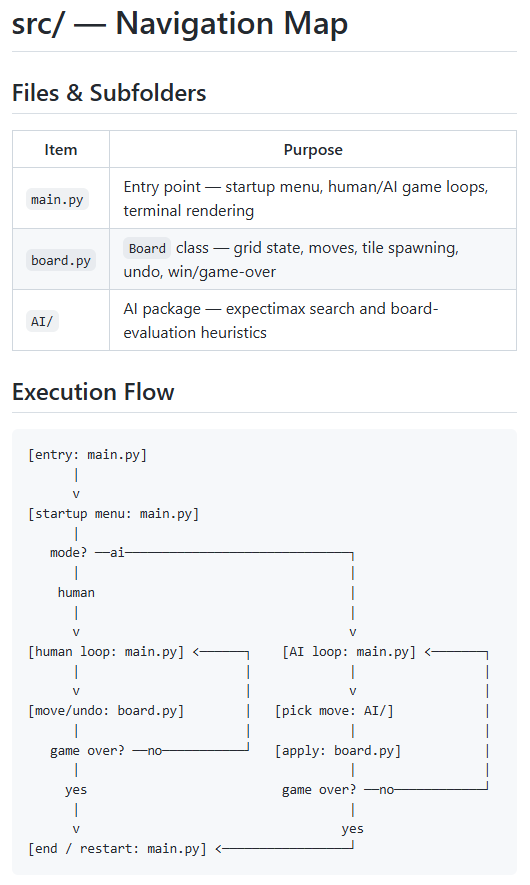}
    \caption{``\lstinline{src/navigation.md}'' in \autoref{lst:repo}.}
    \label{fig:navigation}
\end{figure}

\autoref{fig:navigation} shows the content of ``\lstinline{navigation.md}'' under the subfolder ``\lstinline{src/}'' in the example repository from \autoref{lst:repo}. A navigation file consists of two parts: a file list and an execution flow.
The first part enumerates all files and subfolders in the current directory, each accompanied by a brief description of its functionality. This enables readers to quickly form a high-level understanding of the role of each file.
The second part presents the execution flow across files and subfolders, allowing readers to grasp the overall program structure and locate relevant code. In \autoref{fig:navigation}, execution begins at the entry point in ``\lstinline{main.py}'', where the program first checks whether the user has selected AI mode or human mode to play the game. If AI mode is selected, the program invokes code in the subfolder ``\lstinline{AI/}'' to run the automatic game-playing strategy powered by AI algorithms. In both modes, board-related logic is handled by ``\lstinline{board.py}''. For example, if a user is interested in the detailed logic of human mode, we could proceed directly to the source-code file ``\lstinline{main.py}''.

\subsection{Writing Code Files}
\begin{figure}
    \centering
    \includegraphics[width=0.99\linewidth]{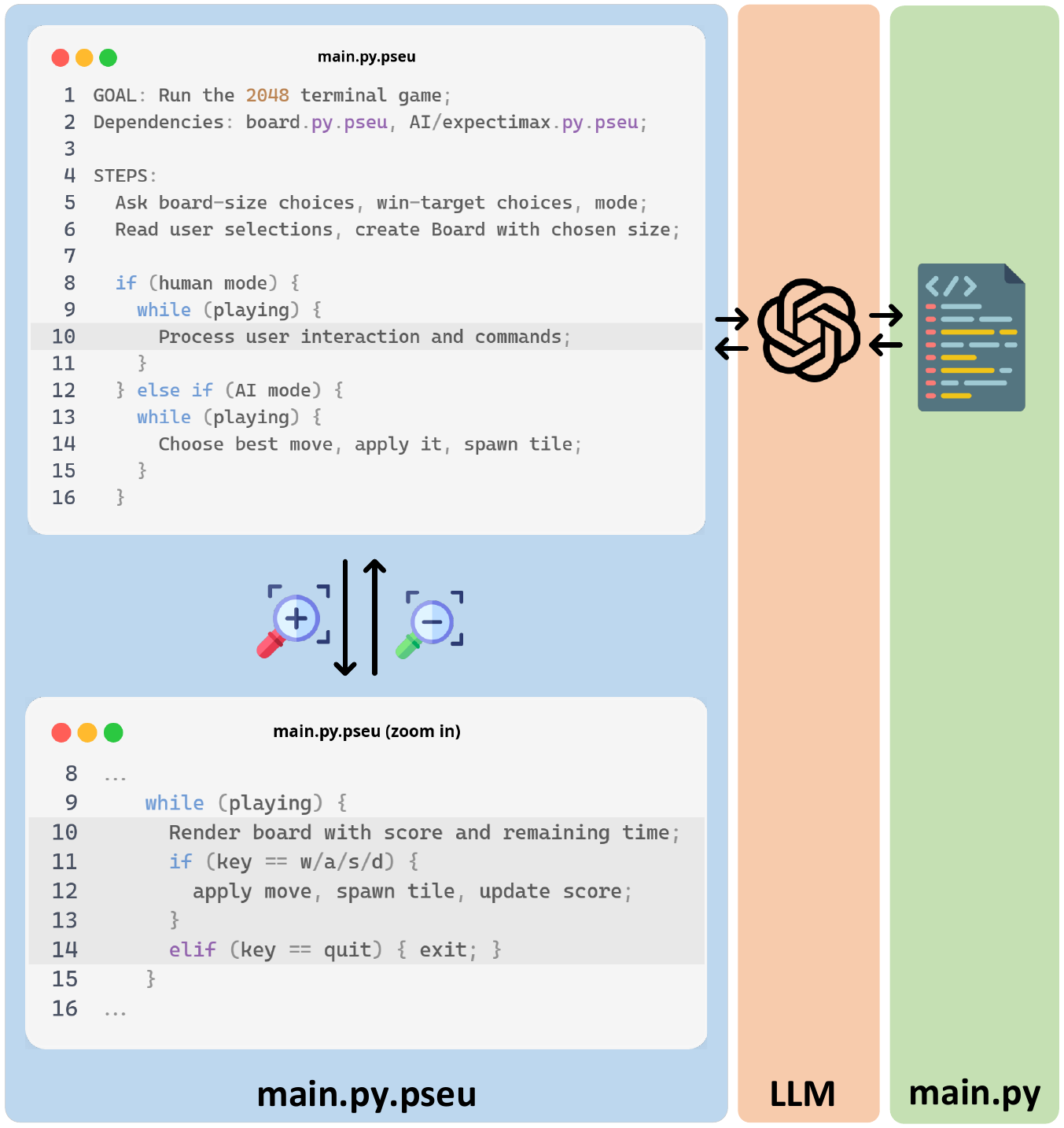}
    \caption{The bidirectional relationship between ``\lstinline{main.py}'' and its pseudocode ``\lstinline{main.py.pseu}'' from \autoref{lst:repo}. An LLM facilitates seamless conversion between the source code and pseudocode. Developers interact with the system by zooming in/out and directly modifying the pseudocode.}
    \label{fig:file}
\end{figure}

After locating the target source code file through the navigation files, we open the corresponding pseudocode file and perform inspection and revision. This process involves four stages: converting source code into pseudocode, semantic zooming on the pseudocode, revising the pseudocode, and synchronizing changes from the pseudocode back to the source code file. In our example, we open ``\lstinline{main.py.pseu}'', as shown in \autoref{fig:file}.

\paragraph{Pseudocode generation}
Each pseudocode representation is generated by \method from its corresponding code file. To support semantic zooming over the pseudocode, we define a formal grammar for the representation, as described in \autoref{sec:pseudocode}. The translation is performed by LLM-based coding agents, such as Claude Code and OpenAI Codex. Specifically, the prompt provides the pseudocode grammar together with an example. In \autoref{fig:file}, the generated pseudocode is shown in the top-left corner.

\paragraph{Semantic zooming}
Given the pseudocode, users expand or collapse specific portions to focus on the semantic logic most relevant to their goals if necessary. This step is also done by LLM-based coding agents, allowing users to provide prompts flexibly. In \autoref{fig:file}, suppose we are curious about what algorithm is implemented for the AI mode, we could zoom in on line 10. Specifically, users could provide the prompt: ``\lstinline{Clarify the AI mode in line 10 at main.py.pseu}'' to \method. Then line 10 in the pseudocode is expanded to lines 10--14 in the bottom-left corner code block. Users could further adjust the abstraction level of the pseudocode until the semantic logic most relevant to their goals is clear enough.

\paragraph{Pseudocode Revision}
Given the adjusted pseudocode, users may revise it to achieve their goals. This step can be performed either manually or by LLM-based coding agents. It is the core user-interaction step in \method and the most significant distinction from the user interface of pure LLM-based coding agents. For example, in \autoref{fig:file}, suppose we would like to add a next move suggestion for human mode, we could manually insert one line ``suggest next move based on a simple heuristic algorithm'' before line 11 in the ``\lstinline{main.py.pseu (zoom in)}'' in the bottom-left corner of \autoref{fig:file}. Alternatively, we could instruct an LLM to perform the revision using the prompt: ``\lstinline{I want to use a simple heuristic algorithm to suggest next move for human mode in main.py.pseu}.'' 

\paragraph{Pseudocode-code sync}
After the pseudocode is modified, the changes are propagated back to the source code, and the pseudocode is also re-generated from the revised source code. Both steps are done by LLM-based coding agents automatically. 
To record the changes in the pseudocode, we create a backup file for each pseudocode file. When the pseudocode file is modified, we can compare it to the backup to observe the modified context and pass it to the LLM-based agents. 
In \autoref{fig:file}, \method maintains a backup file, `\lstinline{.main.py.pseu}'', which does not include lines 10--14 of `\lstinline{main.py.pseu}''. \method then compares the differences between the two files and uses them to guide LLMs in revising the corresponding source code file. Afterwards, \method regenerates the pseudocode from the revised source code to reflect the latest version of the program. This is particularly useful when LLMs do not correctly update the source code according to the pseudocode changes, as users can further inspect the regenerated pseudocode and perform another round of revision.
We argue that \method provides an efficient communication channel between humans and LLMs, enabling a controllable and effective refinement process for code writing using LLMs.

\subsection{Implementation}

We implemented \method as a set of skills. Skills are pre-packaged folders that contain best practices and task-specific instructions for LLM-based agents. Each skill includes a \lstinline{SKILL.md} file that stores condensed guidance accumulated through iterative experimentation, helping LLMs produce higher-quality outputs for specific tasks. Skills are widely supported by existing LLM-based coding agents, such as Claude Code and Codex. We developed our implementation on top of Claude Code, one of the most popular coding agents. Because the skill interface is portable, we believe \method can be adapted by other agent frameworks. %Specifically, we implemented the following skills.

%\begin{itemize}
%    \item \textbf{navigation \textless{}path\textgreater{}}: Generate navigation files for the specified folder and all its subfolders.
%    \item \textbf{code2pseu \textless{}code file(s)\textgreater{}}: Generate pseudocode representations for the given source code files.
%    \item \textbf{pseu2code \textless{}code file(s)\textgreater{}}: Translate pseudocode representations into the corresponding source code files. This command is typically used after pseudocode edits to revise the original code.
%    \item \textbf{zoom \textless{}pseudocode file\textgreater{} \textless{}prompt\textgreater{}}: Adjust the abstraction level of a pseudocode file according to the prompt.
%    \item \textbf{prompt2pseu \textless{}code file(s)\textgreater{} \textless{}prompt\textgreater{}}: Use a prompt to create or revise pseudocode files. This is particularly useful when starting a new project, where users can quickly draft an initial pseudocode design from natural language instructions.
%\end{itemize}
\section{Pseudocode Grammar}\label{sec:pseudocode}

%To utilize pseudocode as the abstract language for vibe coding, we need to specify a grammar for the pseudocode. 

%Traditional pseudocode is entirely informal. However, to utilize it as a bidirectional bridge between humans and LLMs, we introduce a semi-formal variant. We define a minimal EBNF grammar strictly to formalize the control flow structures (e.g., if, while, for). This provides the LLM with unambiguous architectural boundaries. Meanwhile, the actual execution steps within these structures remain unconstrained, informal natural language strings, preserving the human-readable flexibility of traditional pseudocode.

\subsection{Motivation}
Pseudocode has traditionally been an informal description of program semantics and logic. As a result, there is no universally accepted standard grammar. While this informality benefits human readers, it is insufficient for our setting; the complete lack of grammar introduces architectural ambiguity that LLMs frequently misinterpret during semantic zooming. To resolve this, we adopt a semi-formal approach. We introduce a grammar designed strictly to constrain the program's control flow (\eg, loops and conditionals), providing the LLM with unambiguous architectural boundaries. Meanwhile, the operational steps within these structures remain unconstrained, informal natural language strings. This ensures the pseudocode remains as easy for humans to read as traditional informal styles, while providing enough structural scaffolding to guide and constrain LLM code generation.

%Traditional pseudocode is entirely informal. However, to utilize it as a bidirectional bridge between humans and LLMs, we introduce a semi-formal variant.
%Pseudocode has traditionally been an informal description of program semantics and logic, targeting human-readable explanations of algorithms without the strict syntax rules of a programming language. As a result, there is no universally accepted standard grammar for pseudocode. Existing pseudocode styles typically follow practical guidance for specific requirements, such as teaching and education~\cite{pseudocodeguidance}, without a strict or consistent grammar. Such guidance is insufficient for our setting because the lack of grammar introduces ambiguity, variability, and inconsistencies that humans can often resolve, but LLMs may misinterpret for semantic zooming.

%To address this, we establish two design goals for our pseudocode grammar:
%\begin{itemize}
%\item The pseudocode should minimize the cognitive overhead for humans by keeping its syntax as close as possible to natural language.
%\item The pseudocode should provide sufficient structural information to guide and constrain LLMs in generating correct and faithful code.
%\end{itemize}

\begin{figure}
\vspace{-2mm}
\begin{lstlisting}[caption={The grammar of pseudocode in EBNF.},captionpos=t, language=bash, label=lst:grammar, escapeinside=@@, morekeywords={Pseudocode, Goal, Dependencies, Steps, Statement, SimpleStmt, WhileStmt, IfStmt, ForStmt, Cond, Description}]
Pseudocode ::= Goal Dependencies Steps 
Goal ::= 'GOAL:' Description ';'
Dependencies ::= 'Dependencies:' Description ';'
Steps ::= 'STEPS:' (Statement)+
Statement ::= SimpleStmt
                 | WhileStmt
                 | IfStmt
                 | ForStmt
SimpleStmt ::= Description ';'
WhileStmt ::= 'while' '(' Cond ')' '{' (Statement)+ '}'
IfStmt ::= 'if' '(' Cond ')' '{' (Statement)+ '}' 
        ( 'elif' '(' Cond ')' '{' (Statement)+ '}' )* 
        ('else' '{' (Statement)+ '}')?
ForStmt ::= 'for' '(' Cond ')' 
                   '{' (Statement)+ '}'
Cond ::= [a-zA-Z0-9]+
Description ::= [a-zA-Z0-9]+
\end{lstlisting}
\vspace{-8mm}
\end{figure}

\subsection{Grammar}
To formally specify the grammar, we adopt Extended Backus--Naur Form (EBNF)~\cite{feynman2016ebnf}, which provides a precise yet concise notation for expressing context-free grammars. \autoref{lst:grammar} presents the resulting pseudocode grammar in EBNF.

The pseudocode consists of three major components: \textbf{\small Goal}, \textbf{\small Dependencies}, and \textbf{\small Steps}. \textbf{\small Goal} is a one-sentence summary that captures the high-level semantics of the source code. \textbf{\small Dependencies} is a list of other code files on which the current file depends. It enables users to navigate to related files when the current file does not contain the required semantic logic. \textbf{\small Steps} is an ordered sequence of concrete actions used to realize the goal. 
Each step corresponds to a \textbf{\small Statement}, which may be either a simple natural-language sentence (\ie, \textbf{\small SimpleStmt}) or a control structure (\ie, \textbf{\small WhileStmt}, \textbf{\small IfStmt}, or \textbf{\small ForStmt}). Each control structure contains a \textbf{\small Cond} and a nested list of \textbf{\small Statement} enclosed in brackets \textbf{\{\}}. This indicates that when \textbf{\small Cond} evaluates to true, the enclosed statements are executed. Finally, \textbf{\small Description} denotes a natural-language sentence.

Compared to the pseudocode styles proposed for educational purposes~\cite{pseudocodeguidance}, our grammar is deliberately simpler: it includes only the minimal set of branch and loop control structures, while all other aspects are expressed in natural language. This design offers several advantages. First, it reduces the learning curve for humans, as the pseudocode closely resembles everyday language and avoids additional syntactic burdens such as explicit variable declarations or rigid typing rules. Second, it provides enough structural scaffolding to help LLMs avoid common pitfalls, such as misinterpreting the intended control flow or introducing spurious logic. Third, the separation between the high-level \textbf{\small Goal} and the procedural \textbf{\small Steps} mirrors how developers naturally think about code: they start with an overall objective and then refine it into smaller, executable actions. 
\section{Evaluation Setup}

\subsection{Design}
To evaluate our system, we conducted a user study that employed a within-subjects design with two conditions, \emph{\cc}, which is a source-code interaction with an LLM assistant, and \method, which is our pseudocode-based semantic-zooming workflow, layered on the same LLM assistant. Each participant completed both conditions; order (\cc-first versus \method-first) was counterbalanced across participants. Each participant solved the same two programming tasks under each condition. The primary confirmatory hypothesis and analysis plan were pre-registered prior to data collection. The report is available here: \url{https://aspredicted.org/hi97f8.pdf}.

\paragraph{Baseline system} The control condition used \cc v2.1.84 with Claude Sonnet~4.6, which was the latest default model at the time of the study. The \method condition was built on top of the same version of \cc with the same underlying model, so that the two conditions differ only in the pseudocode interface layer.

\paragraph{Tasks} Participants completed two programming tasks per condition, selected to cover complementary scenarios described in \autoref{sec:approach}. 
\add{To prevent cognitive fatigue from confounding our results~\cite{baum2019associating, fritz2014using}, we limited the study to two tasks per participant. This approach prioritizes precise behavioral observation over broad task generalization, a standard methodological trade-off in empirical software engineering research~\cite{stol2020guidelines}.}
\emph{T1 -- Greenfield implementation} asked participants to implement the sliding-tile puzzle \emph{2048} from scratch, matching the behavior of a provided reference binary; T1 primarily exercises design-time control. \emph{T2 -- Feature request in an existing codebase} asked participants to implement a feature request for the LLM fine-tuning framework InstructLab (PR~\#3286)\footnote{\url{https://github.com/instructlab/instructlab/pull/3286}}: adding default \texttt{student\_model\_id} and \texttt{teacher\_model\_id} fields to the \texttt{general} configuration section and propagating them through data generation and training so they take effect unless overridden via the command line. T2 primarily involved code comprehension and targeted modification in an unfamiliar multi-file repository. A recent study~\cite{chen2026featbench} showed that T2 is challenging, as all existing LLM-based coding agents cannot successfully solve it.

\paragraph{Procedure} Sessions were run in person on identical hardware and were proctored by the research team. Due to a limited pool of \cc accounts, participants were scheduled in pairs per session. Each session lasted $60$ minutes and began with a $5$-minute tool-orthogonal tutorial on the condition to which the participant was first assigned. Participants completed the post-condition portion of the survey immediately after each condition, and a final overall-impressions section after both conditions were complete. Each participant received an honorarium equivalent to approximately $30$~USD.

\subsection{Participants}
32 participants from two local universities signed up for our study, of which $N{=}26$ attended. The mean age was $24.6$ years ($\text{SD} = 2.02$; one participant did not report age). Gender distribution was $14$ male, $10$ female, $1$ non-binary, and $1$ not reported. The participants were all master's students in computer science or a related subject. Study year was approximately balanced (1st year: $9$; 2nd year: $9$; 3rd year or later: $7$; $1$ not reported). Self-reported programming experience was high ($14$ reported $>5$ years; $10$ reported $3$--$5$ years; $1$ reported $1$--$2$ years; $1$ not reported), and LLM-tool usage was frequent ($8$ used them daily, $13$ weekly, $3$ monthly, $1$ never, $1$ not reported). Order assignment was balanced: $13$ participants completed the \cc condition first and $13$ completed the \method condition first.

\subsection{Instruments}
After each condition, participants completed a survey booklet comprising the System Usability Scale \cite{brooke1996sus}; 10 items scored on a 0--100 scale, where scores above $68$ indicate above-average usability and above $80.3$ indicate excellent usability; the NASA Task Load Index (NASA-TLX)~\cite{hart1988development}; six subscales rated 0--100 in steps of $5$; a self-report of code understanding (CU, 7-point Likert); and two code-revision items (CR-Locate and CR-Ease, both 7-point Likert). The \method booklet also contained a 7-point rating of zoom-command helpfulness (CZ-Zoom) and two 7-point agreement items on whether \method improves understanding (CZ-Compare-Understand) and revision (CZ-Compare-Revise) relative to working with source code directly (see Figure~\ref{fig:custom_items}). To contextualize the responses, students were also invited to answer freeform questions on when \method was particularly helpful or unhelpful, and how \method could be improved.

\subsection{Protocol Amendment}
\label{sec:amendment}
After the first week of data collection, we amended the survey in two ways, which affected participants $11$--$26$:

\begin{enumerate}
  \item \textbf{Addition of an actual-learning measure (RQ2, below).} The pre-registered confirmatory measure (CU) is a \emph{self-report} of understanding. We realized that behavioral evidence of code awareness would be a useful complement to this subjective measure, because self-reports can dissociate from actual comprehension. We therefore added two task-specific multiple-choice content questions per condition (``QT1'' and ``QT2''), each with eight answer options, to probe participants' knowledge of the codebase.
  
  \item \textbf{Rewording of the agency/control item.} In the original booklet, the \method-section item read, ``I felt in control of the design decisions when working with \method, as opposed to having Claude make all decisions for me'' (agree/disagree, 7-point). We realized that this framing is biased and unsatisfactory because it presupposes that Claude ``makes all decisions'' for the user, which primes agreement with \method. We replaced it with a bipolar comparison item administered once after both conditions: ``Under which programming mode did you feel more in control of your design decisions?'' anchored by $1 = $ \cc and $7 = $ \method.
\end{enumerate}

These changes are deviations from the pre-registered protocol. We report them here in full and treat all analyses that depend on them (RQ2; the bipolar control item) as exploratory and restricted to participants $11$--$26$ ($n{=}16$). The other pre-registered measures (CU, SUS, NASA-TLX, CR-Locate, CR-Ease, CZ-Zoom, CZ-Compare-Understand, CZ-Compare-Revise) were not affected and use the full sample ($N{=}26$).

\subsection{Research Questions}

Our main research questions were:

\begin{enumerate}[
    label=\textbf{RQ\arabic*},
    leftmargin=*,
    labelindent=\parindent,
    labelsep=1em,
    align=left,
    itemindent=2em
]

\item[\textbf{RQ1}] Do participants using \method report higher self-rated code understanding than when using \cc alone?

\item[\textbf{RQ2}] Do participants using \method show higher actual code awareness, as measured by multiple-choice content questions about the codebase, than when using \cc alone?

\end{enumerate}

RQ1 was pre-registered as the confirmatory hypothesis; RQ2 was added as a secondary, pre-specified hypothesis during a protocol amendment after the first week of data collection, with its measure and analysis plan fixed before any of the $n=16$ amended-protocol data were collected.

We also had exploratory comparisons, which were pre-registered as secondary. We computed paired comparisons of SUS total scores, NASA-TLX and its subscale ratings, CR-Locate, and CR-Ease between conditions. Within the \method condition, we further report CZ-Zoom, CZ-Compare-Understand, and CZ-Compare-Revise (each tested against the neutral midpoint of $4$), and the bipolar control-preference item (tested against $4$, with $1 = $ \cc and $7 = $ \method).

\subsection{Analysis Plan}
The confirmatory test for RQ1 was a paired-samples $t$-test on the code understanding question (CU), with a Wilcoxon signed-rank test substituted if the normality assumption on the difference scores was violated (Shapiro--Wilk, $\alpha = .05$). The same rule was applied to all exploratory paired comparisons. Effect sizes are reported as Cohen's $d$ for $t$-tests and the matched-pairs rank-biserial correlation $r_{rb}$ for Wilcoxon tests. For single-sample items, we report one-sample $t$-tests against the scale midpoint ($4$) with Wilcoxon as a robustness check. Per the pre-registered exclusion rule, participants who failed to complete both conditions were excluded; none of the analyzed $26$ had missing data on the pre-registered primary outcomes. All tests are two-sided; we do not adjust for multiple comparisons given the exploratory framing of the secondary analyses, but readers should interpret secondary $p$-values with that caveat in mind. Analyses were performed in Python 3.14 using \texttt{pandas} and \texttt{scipy.stats}.

\section{Results}

\subsection{Self-Reported Code Understanding (RQ1)}

Participants rated their code understanding substantially higher in the \method condition ($M = 5.50$, $SD = 1.10$) than in the \cc condition ($M = 3.35$, $SD = 1.79$), a mean difference of $2.15$ scale points ($SD_{\text{diff}} = 2.11$; $Mdn_{\text{diff}} = 2$). Twenty of $26$ participants rated \method strictly higher than \cc on this item; $4$ rated the two equal and $2$ rated \cc higher. Shapiro--Wilk on the difference scores was not significant ($W = 0.93$, $p = .081$), so the pre-registered paired-samples $t$-test was used as the primary analysis. It showed a large and highly reliable effect in favour of \method, $t(25) = 5.20$, $p < .001$, $d = 1.02$. A Wilcoxon signed-rank test, reported for robustness, reached the same conclusion, $W = 9.00$, $p < .001$, $r_{rb} = 0.93$. \textbf{RQ1 is supported.}

\subsection{Actual Code Awareness (RQ2)}
Among the $16$ participants (IDs $11$--$26$) who completed the added content questions, correct-response rates were markedly higher in the \method condition than in the \cc condition on both items. On QT1 (algorithm used by the AI mode), accuracy was $62.5\%$ after \method and $6.25\%$ after \cc; on QT2 (fallback source of the student model ID), accuracy was $50.0\%$ after \method and $12.5\%$ after \cc. Aggregated, participants answered $18$ of $32$ content questions correctly in \method versus $3$ of $32$ in \cc. Analyzed as per-participant sums ($0$--$2$), the mean score was $1.13$ ($SD = 0.81$) for \method and $0.19$ ($SD = 0.40$) for \cc, a mean difference of $0.94$ ($SD_{\text{diff}} = 0.93$). Shapiro--Wilk on the difference was significant ($W = 0.87$, $p = .028$), so a Wilcoxon signed-rank test was used as the primary analysis: $W = 4.00$, $p = .005$, $r_{rb} = 0.90$. The paired $t$-test agreed, $t(15) = 4.04$, $p = .001$, $d = 1.01$.

\subsection{Exploratory: SUS, NASA-TLX, and Code Revision}
Table~\ref{tab:exploratory} summarizes the exploratory paired comparisons. There was no reliable difference in SUS total scores (\cc $M = 71.25$, \method $M = 69.04$; $t(25) = -0.50$, $p = .62$), and the point estimates suggest that participants perceived the two environments to be similarly usable. On the NASA-TLX, the \method condition was rated as requiring more mental effort and greater overall effort: Mental Demand, Wilcoxon $W = 89.0$, $p = .047$; Effort, $t(25) = 2.97$, $p = .006$, $d = 0.58$. Physical Demand was also rated higher in \method (Wilcoxon $W = 16.5$, $p = .042$, $r_{rb} = 0.64$), though absolute levels on this subscale were low in both conditions ($M$s $\leq 17$). Temporal Demand, Performance, and Frustration did not differ. The aggregate NASA-TLX score was numerically higher in \method but not statistically different ($M = 32.1$ versus $27.7$, $p = .21$, $d = 0.25$). For the code-revision items, CR-Locate was higher in \method ($M = 5.04$) than in \cc ($M = 4.27$), but with a difference that was not significant by both the pre-registered Wilcoxon (normality violated, $W_{\text{Shapiro}} = 0.90$, $p = .013$; Wilcoxon $W = 32.0$, $p = .057$) and the $t$-test ($t(25) = 2.02$, $p = .055$). CR-Ease did not differ between conditions ($M = 5.15$ vs.\ $5.35$; Wilcoxon $p = .48$).

\begin{table}[t]
\centering
\caption{Exploratory paired comparisons, \method (CZ) versus \cc (CC) ($N{=}26$). The $p$-values were non-significant ($> 0.05$).}
\begin{tabular}{l S[table-format=2.2] S[table-format=2.2] S[table-format=1.3] S[table-format=+1.2]}
\toprule
Measure & {CC $M$} & {CZ $M$} & {$p$} & {Effect} \\
\midrule
SUS            & 71.3 & 69.0 & 0.625 & -0.10 \\
TLX: Aggregate & 27.7 & 32.1 & 0.211 & +0.25 \\
CR-Locate      &  4.3 &  5.0 & 0.057 & +0.53 \\
CR-Ease        &  5.2 &  5.4 & 0.480 & +0.19 \\
\end{tabular}
\vspace{-4mm}
\label{tab:exploratory}
\end{table}

\begin{figure*}[t]
\centering
\small
\renewcommand{\arraystretch}{1.15}
\begin{tabularx}{\textwidth}{@{}l >{\raggedright\arraybackslash}X@{}}
\toprule
\textbf{ID} & \textbf{Question} \\
\midrule
\multicolumn{2}{@{}l}{\textit{Per-condition self-reports}}\\
CU        & I was able to understand the overall structure and logic of the codebase with confidence using [CC / CZ]. \\
CR-Locate & I was able to locate the relevant code sections efficiently using [CC / CZ]. \\
CR-Ease   & It was easy to make the necessary code changes using [CC / CZ]. \\
\addlinespace
\multicolumn{2}{@{}l}{\textit{CZ-specific items}}\\
CZ-Zoom              & I felt the \emph{zoom} command was helpful for understanding the codebase at different levels of abstraction.\\
CZ-Compare-Understand & Compared with directly reading source code, CodeZoom makes it easier for developers to understand the codebase. \\
CZ-Compare-Revise    & Compared with manually revising source code, CodeZoom facilitates more efficient localization and modification of relevant code sections.\\
\addlinespace
\multicolumn{2}{@{}l}{\textit{Agency/control item}}\\
Control (original, P1--10)  & I felt in control of the design decisions when working with CodeZoom, as opposed to having Claude make all decisions for me. \\
Control (revised,  P11--26) & Under which programming mode did you feel more in control of your design decisions? \\
\addlinespace
\multicolumn{2}{@{}l}{\textit{Actual-learning items}}\\
QT1 & What algorithm is used in your implementation for the AI mode of 2048? \\
QT2 & Before modification, if the student model ID is not provided via the CLI during data generation, where is the student model retrieved from?\\
\end{tabularx}
\caption{Custom (non-standardised) survey items used in the study, in addition to the SUS~\cite{brooke1996sus} and NASA-TLX~\cite{hart1988development}.}
\label{fig:custom_items}
\vspace{-4mm}
\end{figure*}

\subsection{Exploratory: \method-Specific Items}
All three \method-specific items were rated well above the neutral scale midpoint of $4$ (Table~\ref{tab:czitems}). Participants found the zoom command helpful for understanding the codebase at different levels of abstraction ($M = 5.23$, $SD = 1.27$; $t(25) = 4.92$, $p < .001$). They also agreed strongly that, compared with directly reading source code, \method makes it easier to understand the codebase ($M = 5.65$, $SD = 0.94$; $t(25) = 9.01$, $p < .001$), and that, compared with manually revising source code, \method facilitates more efficient localisation and modification of relevant sections ($M = 5.58$, $SD = 1.03$; $t(25) = 7.83$, $p < .001$).

\begin{table}[t]
\centering
\caption{\method-specific items, tested against the neutral midpoint of $4$ on a 7-point scale ($N{=}26$). The $p$ column reports the paired one-sample $t$-test; Wilcoxon signed-rank gives $p < .001$ for each item as well.}
\begin{tabular}{l S[table-format=1.2] S[table-format=1.2] S[table-format=1.3]}
\toprule
Item & {$M$} & {$SD$} & {$p$} \\
\midrule
CZ-Zoom               & 5.23 & 1.27 & {$<.001$} \\
CZ-Compare-Understand & 5.65 & 0.94 & {$<.001$} \\
CZ-Compare-Revise     & 5.58 & 1.03 & {$<.001$} \\
\end{tabular}
\label{tab:czitems}
\vspace{-5mm}
\end{table}

\subsection{Exploratory: Agency/Control Preference}
As noted in Section~\ref{sec:amendment}, the original CZ-Control item is not pooled with the bipolar version because they measure different constructs. For completeness, we report both.

For the de-biased \emph{bipolar} item administered to participants $11$--$26$ ($n{=}16$; $1 = $ \cc, $4 = $ neutral, $7 = $ \method), mean preference was $5.38$ ($SD = 1.41$, $Mdn = 5$), significantly above the neutral midpoint, $t(15) = 3.91$, $p = .001$; Wilcoxon $W = 14.5$, $p = .005$. A count-based summary is informative: $15$ of $16$ participants expressed a preference toward \method, $0$ were neutral, and $1$ participant expressed a preference toward \cc (rating $= 1$). For the \emph{original} \method-only biased item administered to participants $1$--$10$, responses were $9\times 6$ and $1\times 5$ ($M = 5.90$, $SD = 0.32$; range $5$--$6$), consistent with the interpretive concern that the item's framing suppressed disagreement.

\subsection{Exploratory: Task Performance}
For completeness, we also analyzed task performance. We include task performance only on the second task (T2), since T1 lacks an automatic oracle for comparing a participant's from-scratch 2048 implementation against the reference binary, whereas T2 provides ground truth via the corresponding pull report (PR \#3286) and the project's existing test suite. On T2, success rates were higher with \method ($M = 0.89$, $SD = 0.32$) than with \cc ($M = 0.59$, $SD = 0.50$). Because the paired differences departed significantly from normality (Shapiro-Wilk $W = 0.759$, $p < .001$), we used a Wilcoxon signed-rank test, which confirmed that participants performed reliably better with \method than with \cc ($W = 13.00$, $z = -2.040$, $p = .024$). A common failure mode of \cc was modifying the wrong files. InstructLab's codebase contains LLM model-specific configuration layers that obscure the location of the InstructLab configuration, making it difficult to localize the files relevant to the task requirements. Additionally, we also observed several instances in which \cc attempted to modify test files rather than the corresponding implementation files.

\subsection{Exploratory: Freeform Feedback}
\add{
The free-form responses were broadly positive about \method's core idea; roughly 21 of 26 respondents expressed a favorable overall view. The clearest consensus forming around the code-to-pseudocode direction as an aid to comprehension (cited by 15 of 26). Participants repeatedly described the pseudocode as a structural overview that made unfamiliar code faster to grasp; one called it ``a clear structure or 'map' so you can pinpoint specific information more easily,'' another noted that ``the pseudocode in front of me was particularly useful throughout as it made comprehension of the same code base super fast,'' and several reported that translating code back to pseudocode helped them locate bugs more easily than inspecting model output directly. The most consistent criticism concerned the revision experience (raised by about a third of respondents, or 9 of 26): editing pseudocode by hand felt heavier than conversing with the model, as one participant put it, ``changing pseucode by hand is cumbersome compared to Claude's fast iteration... I think if zoom also was helpful to edit pseucode it would benefit a lot.'' Relatedly, the most common improvement requests were for tighter prompt-driven editing of pseudocode, automatic synchronization between pseudocode and code, and more process feedback while commands run. A minority were less convinced of the abstraction itself (2–3 respondents); one preferred reading the code directly, and another preferred a specification-based intermediate layer over a pseudocode-based one.
}

\subsection{Sensitivity Analysis}\label{sec:sensitivity}
As our study was within-subjects and is therefore susceptible to learning or priming effects, we ran a sensitivity check on RQ1 and RQ2, using between-subjects comparisons restricted to each participant's first condition only: the $13$ participants who saw \cc first contribute their \cc responses, the $13$ who saw \method first contribute their \method responses, and no participant contributes to both arms. Both effects survive at $p<.05$. For RQ1 (self-rated code understanding), Welch's $t(22.8) = 3.36$, $p = .003$, Cohen's $d = 1.32$ ($M_{\cc} = 3.54$, $M_{\method} = 5.69$); the rank-based Mann--Whitney test agrees, $U = 141.5$, $p = .003$. For RQ2 (actual code-awareness score, restricted to participants $11$--$26$ for whom the content questions were administered, $n = 8$ per arm), Welch's $t(10.3) = 4.02$, $p = .002$, Cohen's $d = 2.01$ ($M_{\cc} = 0.13$, $M_{\method} = 1.25$ out of $2$; Mann--Whitney $U = 57.5$, $p = .004$). The point estimates of effect size are larger between-subjects than within-subjects, despite the smaller and unpaired sample. We take this as evidence that the primary findings are not artifacts of within-subject carryover.
\section{Discussion}

\noindent
\emph{\textbf{Interpretation.}}
In our study, we compared \method against a programming environment where the participants had unrestricted access to AI support. Our two primary measures, namely self-rated understanding (RQ1) and objective code awareness (RQ2), were strongly in favor of \method ($d \approx 1.0$, $p < 0.05$). Furthermore, $15/16$ participants preferred \method on the de-biased agency item. Against that, SUS was indistinguishable between conditions ($p=.62$), and subscales of the NASA-TLX showed that \method required significantly more mental effort (Mental Demand $p=.047$; Effort $d=0.58$, $p=.006$).
Our results lend some support to the thesis that \method can help mitigate risks that are sometimes introduced by AI-augmented programming workflows, namely that developers can lose agency of their work and lack an understanding of their own code. At the same time, this cost did not translate into worse coding outcomes: participants in fact performed materially better with \method than in the unrestricted-AI baseline. Our data provides some indication that at least for some programming tasks, \method can help the developer recover the grasp of their own code without sacrificing the productivity gains that motivate AI-augmented workflows in the first place.

\noindent
\add{
\emph{\textbf{Zooming Behavior.}} While subjective ratings indicate that participants found the zoom command helpful ($M=5.23$), we further analyzed their interaction patterns. Every participant executed at least one zoom-in operation. We observed that participants frequently requested to expand specific lines of pseudocode immediately after generating it from the source code. For example, in T1 (2048), after implementing modifications, participants regenerated the pseudocode and zoomed in on the AI execution block to inspect how the next move was calculated. This enabled them to confirm algorithmic choices (\eg, expectimax) that may otherwise be obfuscated in the initial coarse pseudocode. Conversely, this zoom-out behavior frequently occurred after multiple consecutive zoom-in operations to effectively reset the cognitive context. These behavioral patterns confirm that semantic zooming was utilized not merely as a static reading aid, but as an active, dynamic tool to negotiate the abstraction level required for specific comprehension and modification tasks.

%\noindent
%\emph{\textbf{Zooming Stability.}} To evaluate the stability and generalizability of the zooming mechanism, we collected 40 code files from GitHub spanning 13 different programming languages. We applied \method to generate pseudocode representations for each file, executing zoom-in and zoom-out operations to create a three-layered pseudocode hierarchy. We then prompted the LLM to verify whether each layer remained semantically equivalent to the original source code. To mitigate the risk of LLM hallucination, one of the authors manually audited the semantics. Specifically, [XXXX]. Our evaluation confirmed that all generated pseudocode representations faithfully reflected the underlying semantics of the source code across every level of abstraction.
}

\noindent
\emph{\textbf{Scope.}}
\method is designed to improve the controllability of code generation by providing a new communication interface between humans and LLMs, complementing rather than replacing existing LLM-based coding agents. Our goal is to give developers tighter control over the semantic logic of the generated program so that it faithfully follows their intent. Accordingly, we target experienced programmers who already have a clear picture of how their code should be structured, rather than non-experts who rely on LLMs to propose solutions for coding, such as fixing language-level syntax errors or simple bugs. For such experienced developers, \method offers a structured medium in which to specify semantics precisely, enabling LLMs to generate more accurate code with less back-and-forth. We therefore view \method as a layer that augments agents such as Claude Code, and we expect practitioners to continue relying on the agent's native capabilities for issues that lie below the semantic layer \method captures, such as build-error diagnostics, test-driven iteration, and runtime debugging.

\noindent
\emph{\textbf{Necessity.}}
\method remains necessary even as LLMs become more capable. \method targets \emph{controllability}: how developers communicate with LLM agents, rather than the models' raw coding ability. These two dimensions are largely orthogonal. Even a model with perfect coding skills cannot infer a developer's intent beyond what is explicitly communicated. However, natural language is inherently underspecified for program semantics: the same English instruction may correspond to many valid implementations that differ in control flow, data structures, or edge-case behavior. Stronger models may generate more plausible code from vague prompts, but plausibility does not guarantee fidelity to intent, and this gap becomes harder to detect when outputs are consistently fluent. \method addresses this challenge by providing developers with a structured medium to express intent precisely and to validate the generated code against it. If anything, more capable models further increase its value: they can follow precise specifications more faithfully, while also scaling the risk of confidently incorrect code from underspecified prompts.

\noindent
\emph{\textbf{Throughput.}}
The main performance overhead of \method is the latency of the LLM round-trip: every zoom, revision, and code--pseudocode synchronization requires at least one call to the underlying agent. Currently, each call passes the relevant source file together with its pseudocode counterpart, which is acceptable for the single-file or few-file programs used in our study. \method typical responses complete within a few seconds, but will not scale naively to repositories of thousands of files. Reducing this cost will require smarter prompt construction, such as selectively including only the portions of the repository that the navigation file identifies as relevant to the current edit, and caching pseudocode across calls so that unchanged files are not re-translated. %We view this as an engineering challenge rather than a conceptual one: the pseudocode layer itself already provides a natural unit at which to scope prompts.

\noindent
\emph{\textbf{Threats to validity.}}
Our study has some limitations. The within-subjects design could cause carryover effects, since participants solved the same tasks under both conditions; counterbalancing and the between-subjects sensitivity analysis (Section~\ref{sec:sensitivity}) mitigate this to some extent. A single 60-minute lab session with a tool built by the authors could have introduced novelty effects and acquiescence bias. Our study was conducted with master’s students from two universities in the same geographic region. The results may differ for more experienced developers, who may exhibit different work pacing, interruption patterns, or toolchain integration practices. The \method system might be less effective for different tasks; we do not claim that it is a general-purpose solution for all software development tasks, but rather that pseudocode-based semantic zooming is a promising design point for tasks centered on code comprehension and targeted modification; we consider our two tasks to be representative instances of that. We also note that both conditions used a single underlying model (Claude Sonnet~4.6) and \cc version; the size of \method's benefit may shift with future models.
\section{Related Work}

\paragraph{From code to pseudocode}
A complementary research direction is translating source code into pseudocode to support code comprehension. Early work in this area primarily relied on rule-based methods. For example, Sridhara \etal analyzed function definitions using predefined heuristics to generate summary comments~\cite{sridhara2010towards, sridhara2011automatically}, Buse \etal designed heuristic rules to summarize execution specifications~\cite{buse2008automatic}, and Moreno \etal focused on summarizing class definitions with similar techniques~\cite{moreno2013automatic}. While these approaches can produce accurate summaries when their rules align well with the input, they demand significant manual effort to design and are difficult to generalize across programming languages and projects. More recently, with the advent of LLM, data-driven methods have been proposed to automatically convert source code into pseudocode. Oda \etal applied statistical machine translation techniques to this task~\cite{oda2015learning}, while Alokla \etal fine-tuned BART~\cite{lewis2019bart} and proposed a retrieval-augmented transformer for pseudocode generation~\cite{alokla2022pseudocode, alokla2022retrieval}. Other works demonstrate the utility of pseudocode in educational contexts~\cite{kazemitabaar2024codeaid}, or directly leverage LLMs for generation~\cite{gad2022dlbt, yang2021fine}. Building on these efforts, \method exploits the generative capability of LLMs to directly translate source code into pseudocode, while introducing semantic zooming to improve comprehension.

\paragraph{From pseudocode to code}
Similarly, recent research has explored diverse approaches for translating pseudocode into executable source code. Search-based methods such as SPoC leverage compiler feedback to iteratively refine candidate programs~\cite{kulal2019spoc}. Machine learning approaches have also been widely studied: Acharjee \etal employ a sequence-to-sequence model for pseudocode-to-code translation~\cite{acharjee2022seq2seq}. More recently, transformer-based methods have gained traction, with C3PO introducing a lightweight Copy–Generate–Combine mechanism~\cite{c3po2022}, and retrieval-based transformers enhancing translation through external memory~\cite{alokla2022retrieval}. Complementary efforts include survey work highlighting transformer-based translation of pseudocode into Python~\cite{kumar2023review}. Coladder~\cite{yen2024coladder} establishes a unidirectional, multi-layer translation from prompts to source code. In this work, we establish a bidirectional translation between pseudocode and executable source code for efficient control of the code.

\paragraph{LLM for software engineering}
A relevant and broader research direction is leveraging large language models (LLMs) to address problems in software engineering. A recent survey reports that over one hundred papers have already been published in this area~\cite{liu2024large}. For example, Elicitation~\cite{ataei2025elicitron} and SpecGen~\cite{ma2024specgen} employ LLMs to mine user requirements as comprehensively as possible. CodePlan~\cite{bairi2024codeplan} and AgentCoder~\cite{huang2023agentcoder} adopt multi-agent systems to decompose user requirements into sub-tasks for code generation. 
%Testing has also emerged as a key domain where LLMs show promise: BugScope~\cite{guo2025bugscope} applies LLMs for static bug detection, KernelGPT~\cite{yang2025kernelgpt} generates test cases for operating system kernels, and Fuzz4all~\cite{xia2024fuzz4all} creates grammatically correct test inputs for diverse systems. 
Furthermore, SWE-bench~\cite{jimenez2023swe} has become a widely used benchmark for evaluating LLMs on real-world bug fixing. In contrast to these works, our work focus on the controllability of code writing.
\section{Conclusion}
In this paper, we introduced \emph{Code Semantic Zooming}, a paradigm for LLM-assisted code writing that uses pseudocode as a zoomable intermediate layer between developers and coding agents. Our approach pairs a deliberately minimal EBNF grammar for pseudocode with a bidirectional translation to source code, and adds a semantic zoom operation that lets developers dynamically adjust the level of abstraction at which they read and edit a program. We realized this concept in \method, implemented as a portable set of skills on top of Claude Code, and evaluated it in a within-subjects user study with 26 graduate students on two complementary tasks: a greenfield 2048 implementation and a feature-addition task in an existing InstructLab codebase. \method matched Claude Code on usability while producing a large effect on code comprehension (Cohen's $d=1.02$; 56\% vs.\ 9\% correct on objective content questions) and a clear preference on design control ($>$90\% preference rate). We argue that this work offers a novel perspective on LLM-assisted coding, paving the way toward a more controllable, comprehensible, and human-in-the-loop paradigm for software development.

\bibliographystyle{plain}
\bibliography{references}

\end{document}